\documentclass[twocolumn,floatfix]{aa}

\usepackage{amssymb}
\usepackage{amsmath}
\usepackage{hyperref}
\usepackage{booktabs,tabularx}
\usepackage[table,xcdraw]{xcolor}
\usepackage[normalem]{ulem}
\usepackage{graphicx}
\usepackage[varg]{txfonts}

\begin{document}

\title{Evidence for multiple types of post-starburst galaxies} 

\author{
Emma W. Nielsen \inst{1,2} \and
Charles L. Steinhardt \inst{2,3} \and
Mathieux Harper \inst{3} \and
Conor McPartland \inst{1,4} \and
Aidan Sedgewick \inst{1,2}
}

\institute{
Niels Bohr Institute, University of Copenhagen, Jagtvej 155 A, 2200 Copenhagen N, Denmark
\and
DARK Cosmology Center, University of Copenhagen, Jagtvej 128, 2200 Copenhagen N, Denmark
\and
Department of Physics and Astronomy, University of Missouri, 701. S. College Ave., Columbia, MO65203, USA
\and
Cosmic Dawn Center (DAWN), University of Copenhagen, Jagtvej 155 A, 2200 Copenhagen N, Denmark
}

\abstract{The quenching mechanisms of galaxies are not yet fully understood, but post-starburst galaxies provide one explanation for the rapid transition between star-forming and quiescent galaxies. At low redshift, it is generally thought that the starburst initiating the post-starburst phase is merger-driven, however, not all post-starburst galaxies show evidence of a merger, and recent studies suggested that post-starburst galaxies may be produced by multiple distinct mechanisms. This study examines whether multiple types of post-starburst galaxies actually exist, that is, whether the properties of post-starburst galaxies are multimodal. We used uniform manifold approximation and projection (UMAP) to cluster post-starburst galaxies based on spectroscopic data. The results suggest that there are three types of post-starburst galaxies that have dissimilar stacked spectral energy distributions and are separated by a combination of their H$\alpha$ and [OII]$\lambda$3727 line strengths with an accuracy of 91\%. A comparison of various galaxy properties (e.g., emission line strengths, mass and age distributions, and morphologies) indicates that the grouping is not just an age sequence, but may be correlated to the merger-histories of the galaxies. It suggested that the three post-starburst galaxy types have different origins, some of which may not be merger-driven, and that all typical galaxies go through the post-starburst phase at turnoff.}


\maketitle

\section{Introduction}
\label{sec:intro}
Post-starburst (PSB) galaxies are a class of galaxies that are selected to have a large population of A stars (traced by H$\delta$), but no O and B stars (traced by H$\alpha$ and [OII]). This abundance of A stars is typically interpreted as indicating that PSB galaxies have experienced a recent secondary burst in star formation, whereas the absence of short-lived O and B stars suggests that they have minimal current star formation. Thus, a natural explanation, and the basis for the term post-starburst, is that PSB galaxies have rapidly quenched following an unsustainable burst of star formation \citep[e.g.,][]{Vergani_2009, Wilkinson_2017, French_2021}.

Because low-redshift ($z<2$) PSB galaxies exhibit an overabundance of galaxy-galaxy mergers, these mergers are a likely candidate for triggering the starbursts that lead to a subsequent PSB phase. A plausible mechanism is that mergers cause dense gas to compress and thereby allow an abrupt increase in star formation, but the detailed mechanisms for producing PSB galaxies are not yet fully understood \citep{Pawlik_2019, French_2021, Li_2023}. 

Moreover, not all PSB galaxies show evidence of a past or ongoing merger \citep{Pawlik_2015, French_2021, Cheng_2024}. The reason might be that the morphological evidence of the merger fades over time \citep{Pawlik_2015, French_2021}, or alternatively, that multiple types of PSB galaxies are mixed together, some that are merger-driven, and some that are not. \citet{Li_2023} reported that significantly more post-merger PSB galaxies quench their star formation outside-in, whereas slightly more nonpost-merger PSB galaxies quench inside-out. Unless some feedback mechanism causes inside-out-quenching galaxies to subsequently quench outside-in, this observation is consistent with the idea that PSB galaxies are produced by multiple distinct mechanisms.

Even if the PSB phase is entirely merger-driven, \citet{Pawlik_2019} proposed three distinct merger-driven mechanisms that might produce dissimilar classes of PSB galaxies. These include a transition from star-forming to quiescent, in which a merger-induced starburst is rapidly quenched; a scenario in which the galaxy resumes its star formation after the PSB phase; and a scenario in which a quiescent galaxy is rejuvenated by a minor merger. Alternatively, \citet{Steinhardt_2025} proposed a secular phase in which galaxies stop forming O and B stars, but continue to form low-mass stars, including A stars. These so-called red star-forming galaxies (RSFGs) would therefore be selected as PSB, even though they are not associated with a starburst or merging. On the other hand, \citet{Wilkinson_2017} suggested that variations in the PSB selection criteria (which affect other parameters such as color, morphology, and the environment) are merely caused by observing the galaxies in different stages of the PSB phase.

If this hypothesis is true and there are populations of PSB galaxies with distinct origins, these populations might have other distinct properties. For example, a rejuvenated galaxy should be as old and as massive as a quiescent galaxy, whereas the mass of a galaxy that transitions from star-forming to quiescent should be similar to that of galaxies at turnoff. Likewise, a secularly evolving RSFG should be less morphologically disturbed than a PSB galaxy whose starburst was initiated by a major merger. 

In this work, observational low-redshift PSB catalogs are used in an attempt to determine whether there are indeed multiple classes of PSB galaxies.  The results indicate that there are three spectroscopic classes of PSB galaxies with dissimilar properties.  Then, other properties are used to determine whether the grouping arises because the galaxies have different origins. 

In Sect. \ref{sec:galaxy_sample} we describe the galaxy sample and data, and in Sect. \ref{sec:identifying_psb_groups} we determine whether there are multiple types of PSB galaxies, that is, whether their properties are multimodal, using machine-learning to cluster PSB galaxies based on spectroscopic data. Various galaxy properties (e.g., emission line strengths, mass and age distributions, and morphologies) are examined in Sect. \ref{sec:psb_properties} to determine the origins of the proposed types of PSB galaxies in Sect. \ref{sec:potential_origins}. Specifically, three hypotheses are evaluated: (1) the grouping could be an evolutionary sequence; (2) the groups could be produced by mergers that affect different types of precursor galaxies; or (3) some galaxies might instead evolve secularly through the PSB phase. The results are discussed in Sect. \ref{sec:discussion}, and we find that this final scenario is most consistent with observations.  These results even allow the possibility that PSB galaxies are a turnoff mechanism from the star-forming main sequence, in which case, all typical galaxies would go through a PSB phase at turnoff. 

\section{Galaxy sample and data}
\label{sec:galaxy_sample}
This work primarily relies on a sample of 2665 PSB galaxies from \citet{Meusinger_2017}. The catalog includes PSB galaxies with a median redshift $z=0.13$ (16\textsuperscript{th} and 84\textsuperscript{th} percentiles of 0.069 to 0.199) that were recorded in Stripe 82 by the Sloan Digital Sky Survey Data Release 7 \citep[SDSS DR7,][]{Abazajian_2009}. The \citet{Meusinger_2017} galaxies were selected to have equivalent widths EW(H$\delta$) $>3$ Å, EW(H$\alpha$) $>-5$ Å and EW([OII]) $>-5$ Å, where negative values indicate emission, and positive values indicate absorption. Thus, the galaxies have strong H$\delta$ absorption and almost no H$\alpha$ or [OII] emission. The use of H$\alpha$ reduces contamination from dust, but may introduce a bias against galaxies with strong narrow-line active galactic nuclei (AGN), strong shocks (which can be expected post-merging), or that are not yet fully quenched \citep{French_2021, Li_2023}. 

The \citet{Meusinger_2017} catalog provides equivalent widths (EWs) of H$\alpha$, H$\delta$, [OII]$\lambda$3727 and [OII]$\lambda$3729. The spectral energy distributions (SEDs) were retrieved from the SDSS DR7 Data Archive Server, and the galaxy images we used were taken by the Panoramic Survey Telescope and Rapid Response System \citep[Pan-STARRS, ][]{chambers2019panstarrs1surveys}. For each galaxy, photometric stellar masses and star formation rates estimated from nebular emission lines were retrieved from the MPA-JHU catalog \citep{Kauffmann_2003, Brinchmann_2004}. Photometric stellar masses, ages, extinction, and metallicities as well as EWs of H$\alpha$, H$\beta$, [NII], and [OIII] were retrieved from the Portsmouth catalog \citep{Maraston_2013}. The MPA-JHU and Portsmouth catalogs include 2647 and 2568 of the \citet{Meusinger_2017} galaxies, respectively. 

\section{Identifying dissimilar types of PSB galaxies}
\label{sec:identifying_psb_groups}
Dissimilar groups of PSB galaxies can be identified by using a dimensionality reduction technique called uniform manifold approximation and projection \citep[UMAP,][]{mcinnes2020umap}. 
UMAP is an unsupervised machine-learning algorithm that embeddes high-dimensional data in a lower-dimensional space (usually 2D) while preserving most of its topological structure. Here, UMAP was provided with EWs of H$\alpha$, H$\delta$, [OII]$\lambda$3727 and [OII]$\lambda$3729 (the strengths of the spectral lines used to select the galaxies) and was given no information about any other galaxy property. 

UMAP first approximates the topological structure of the data by constructing a graph that connects every galaxy to its $k$ nearest neighbors and weights the connections based on how similar the galaxies are \citep{mcinnes2020umap}. Importantly, the weighted graph allows disjoint groups to exist. UMAP then returnes a 2D embedding of the data where similar galaxies are attracted and dissimilar galaxies are repelled. Consequently, the dimensions of the UMAP embedding have no physical meaning; rather, galaxies positioned close to each other should be interpreted as similar, and vice versa \citep{mcinnes2020umap}, but the precise locations are neither meaningful nor static when the data are provided in a different order. 

\begin{figure}
    \centering
    \includegraphics[width=\linewidth]{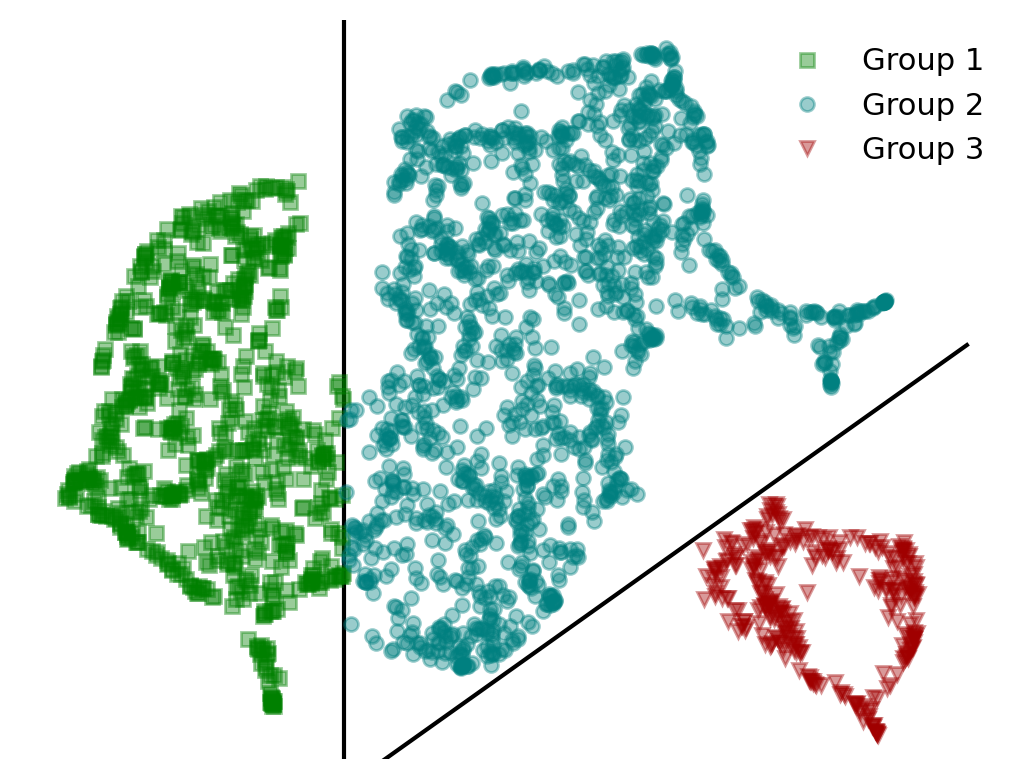}
    \caption{UMAP embedding of the galaxies in the \citet{Meusinger_2017} catalog. Each point represents one galaxy and has an opacity $\alpha=0.4$ to visualize the densities. UMAP was provided with EWs of H$\alpha$, H$\delta$, [OII]$\lambda$3727 and [OII]$\lambda$3729, and default UMAP parameters were used. }
    \label{fig:umap_embedding}
\end{figure}

\begin{figure}
    \centering
    \includegraphics[width=\linewidth]{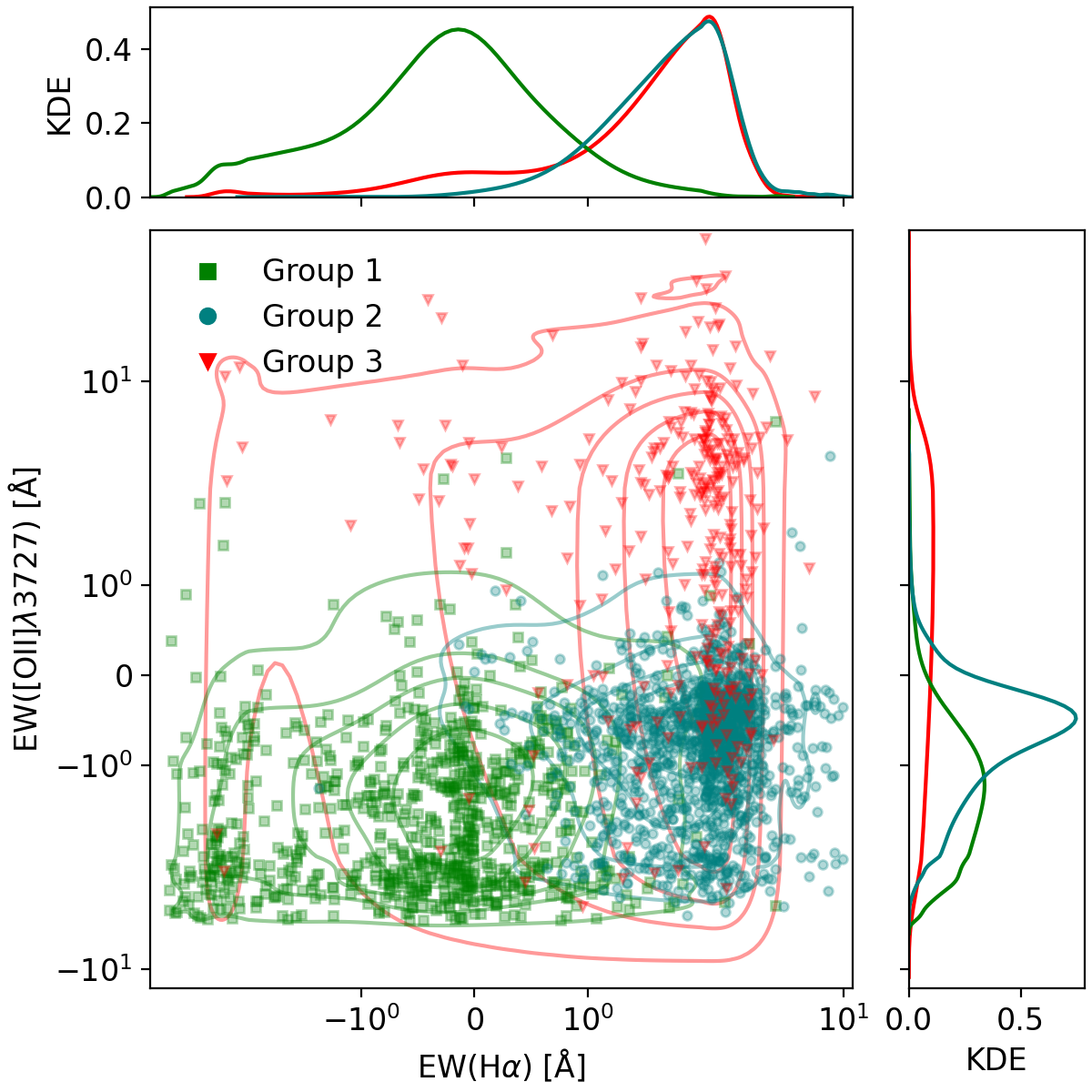}
    \caption{Kernel density estimates (KDEs) of EW(H$\alpha$) vs. EW([OII]$\lambda$3727) for each group. Negative values indicate emission, and the data are shown on a symmetric logarithmic scale. To ensure that the distributions are depicted accurately, the KDE for Group 2 was limited to EW(H$\alpha$)$<10$ Å, including $\geq98\%$ of the galaxies in this group. }
    \label{fig:bimodality_ha}
\end{figure}

Fig. \ref{fig:umap_embedding} shows a UMAP embedding of a single run, where a completely disjoint cluster of galaxies (called Group 3) is identified. Because Group 3 ($N=328$) is completely disjoint, it should be interpreted as a separate group of PSB galaxies with the strong possibility of having a distinct astrophysical origin. Further, Fig. \ref{fig:bimodality_ha} shows a clear bimodality of EW(H$\alpha$), suggesting that the remaining galaxies should be divided into Groups 1 ($N=831$) and 2 ($N=1506$), even though they are only partially separated in the UMAP embedding. A pairwise Kolmogorov-Smirnov (KS) test comparing the univariate distributions of the EWs did not support the null hypothesis that Groups 1 and 2 belong to the same parent population ($p\ll 0.01$). Instead, Groups 1, 2, and 3 can be separated by cuts on their EW(H$\alpha$) and EW([OII]$\lambda$3727) with an accuracy of 91\% (macro-averaged precision and recall are 91\% and 86\%, respectively; see \citealt{Opitz_2024}). Accuracy as a measure of the performance of the separation criteria is biased, however, because the distribution of galaxies between the groups is unbalanced. 

\begin{figure}
    \centering
    \includegraphics[width=\linewidth]{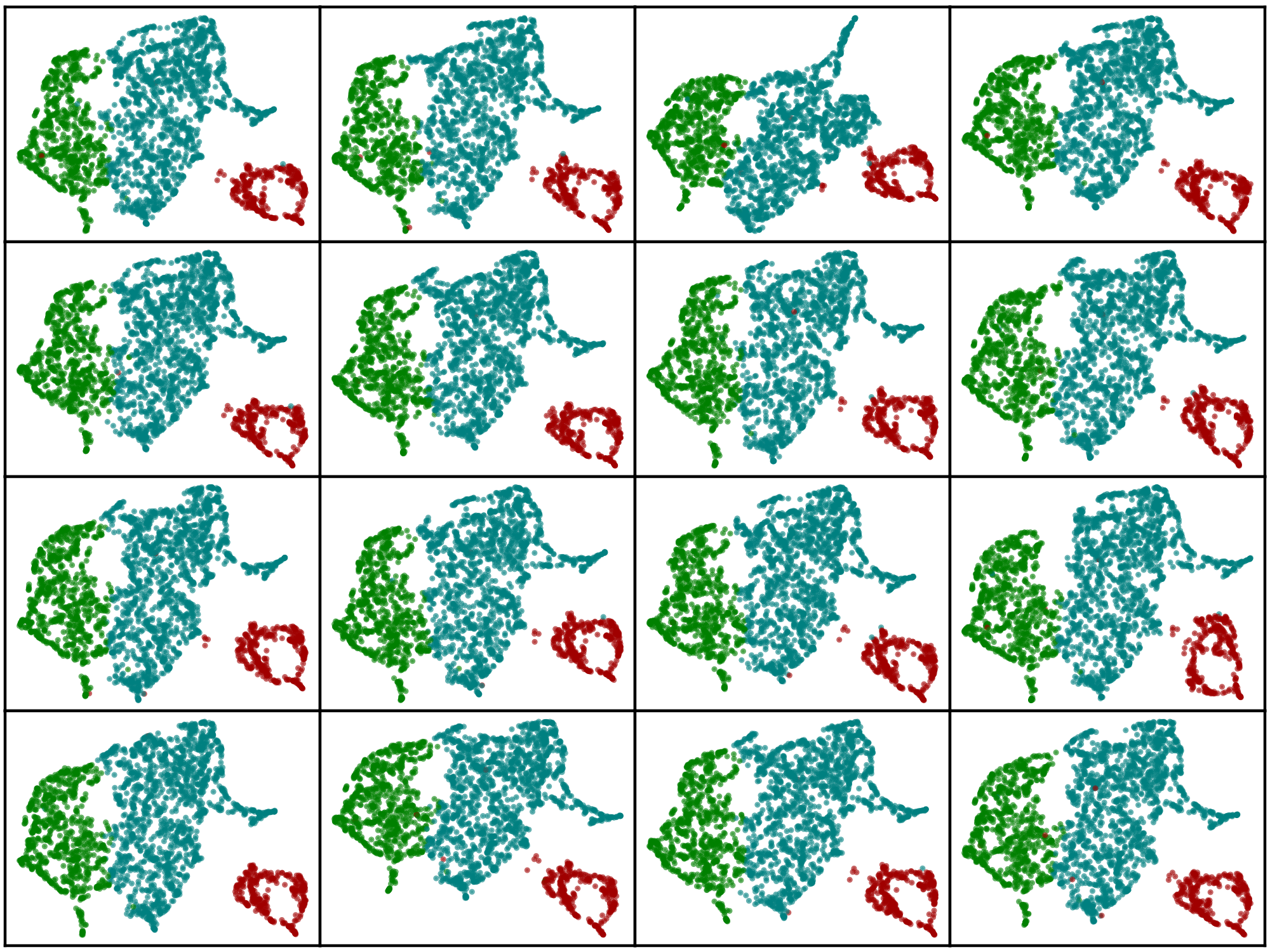}
    \caption{UMAP embeddings of the galaxies in the \citet{Meusinger_2017} catalog sorted randomly into 16 different orders. As in Fig. \ref{fig:umap_embedding}, galaxies in Groups 1, 2, and 3 are shown in green, blue, and red, respectively. }
    \label{fig:umap_16_embeddings}
\end{figure}

It should be noted that UMAP uses stochastic processes when it searches for the $k$ nearest neighbors and the optimal 2D embedding. This results in differences in the 2D embedding from one run to the next \citep{mcinnes2020umap}, with the possibility that individual objects might switch to a different group \citep{Steinhardt2023}. Although the actual locations of the galaxies differ, the structural changes between the 16 randomly selected UMAP embeddings shown in Fig. \ref{fig:umap_16_embeddings} are negligible, and only a few galaxies jump between groups. The separation appears insensitive to random seeds or the choice of the hyperparameter $k$, and instead indicates a true separation into multiple classes. The grouping shown in Fig. \ref{fig:umap_embedding} is used throughout this work.

\section{Properties of PSB galaxy types}
\label{sec:psb_properties}
\subsection{Spectral energy distributions}
Table \ref{tab:group_properties_meusinger} provides an overview of selected spectral properties of the three proposed types of PSB galaxies. Generally, Group 1 emits both H$\alpha$ and [OII], whereas Group 2 emits some [OII], but absorbs H$\alpha$. Group 3 absorbs both H$\alpha$ and [OII]. All groups have strong H$\delta$ absorption. 

\begin{figure*}
    \centering
    \includegraphics[width=.9\textwidth]{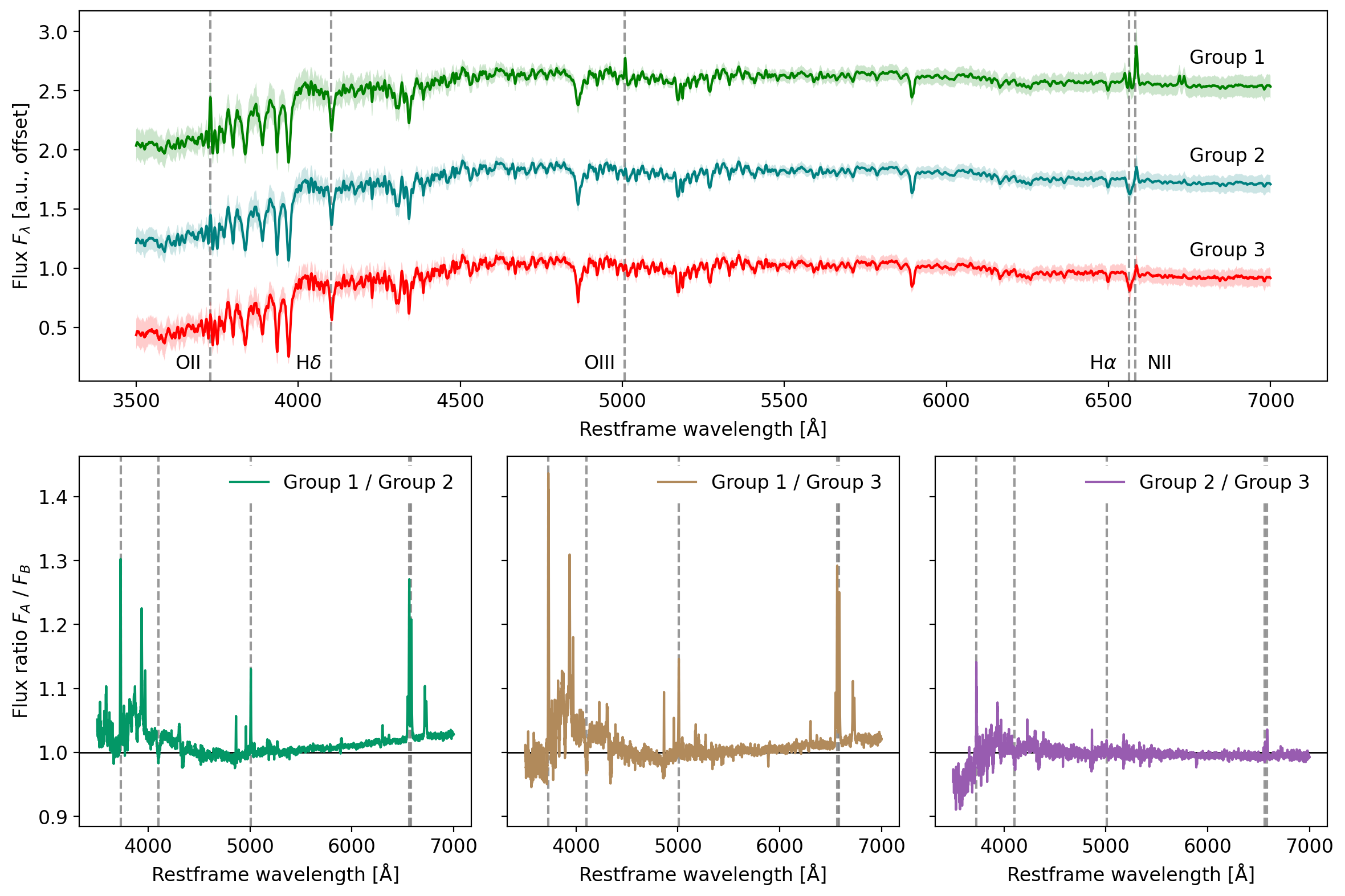}
    \caption{\textit{Top:} Stacked SEDs for the PSB groups identified in the \citet{Meusinger_2017} catalog. Groups 1, 2, and 3 are shown from top to bottom. Prior to stacking, the SEDs were normalized by the mean flux in the 5050-5150 Å region to account for redshift dependences (identical continua were assumed). All SEDs were given equal weight, but only spectra with a signal-to-noise ratio $>5$ for the g, r, and i bands were included (this includes $>97\%$ of the galaxies in the catalog).  16\textsuperscript{th} and 84\textsuperscript{th} percentiles are shown within the shaded area, and selected spectral lines are highlighted. \textit{Bottom:} Flux ratios. }
    \label{fig:stacked_spectra}
\end{figure*}

\begin{figure}
    \centering
    \includegraphics[width=.9\linewidth]{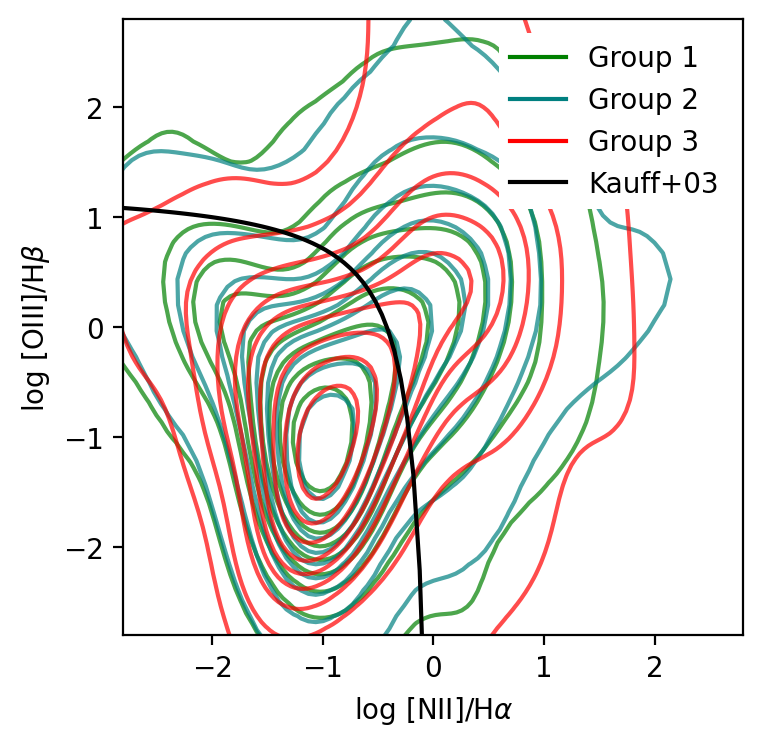}
    \caption{Contour BPT diagrams of the three PSB groups. The EWs of H$\alpha$, H$\beta$, [NII], and [OIII] were retrieved from the Portsmouth catalog. Galaxies above the black line are classified as AGN (see \citet{Kauffmann_2003_agn}). }
    \label{fig:bpt_diagram}
\end{figure}

Fig. \ref{fig:stacked_spectra} shows that Group 1 also emits [OIII] and [NII]. This is indicative of ongoing star formation or AGN activity (possibly initiated by merging) \citep{Baldwin_1981,Kewley_2013}. Given its strong emission lines, Group 1 might contain AGNs, but following \citet{Kauffmann_2003_agn}, many of the galaxies are not classified as AGN. Furthermore, Fig. \ref{fig:bpt_diagram} shows that the groups are not separated in a BPT diagram, indicating that AGN activity does not cause this grouping of PSB galaxies. 

The galaxy spectra shown in the upper panel of Fig. \ref{fig:stacked_spectra} differ primarily in the lines, but the flux ratios (bottom panel) show that Group 1 has greater continuum emission at either end of the visible spectrum. Extinction measurements are consistent with all three groups being identical, however, implying that the enhanced continuum emission toward UV wavelengths has to be explained by something other than dust. For example, Group 1 might contain some massive O and B stars that have not yet died, as also indicated by its H$\alpha$ and [OII] emission. The grouping does not separate AGNs (Fig. \ref{fig:bpt_diagram}), so AGN activity does probably not cause the enhanced continuum emission. Group 3 has greater continuum emission than Group 2 at lower wavelengths.

\begin{table*}
\caption{Median EWs of selected spectral lines for the PSB groups identified in the \citet{Meusinger_2017} catalog.}
\centering
\begin{tabular}{@{}lllll@{}}
\toprule
\textbf{}               & \phantom{$-$}\textbf{Group 1}             & \phantom{$-$}\textbf{Group 2}             & \textbf{Group 3}       \\ \midrule
\textbf{EW(H$\alpha$) [Å]} & $-0.28$ ($-1.91$ to 0.45)                         & \phantom{$-$}2.26 (1.48 to 3.27) & 2.14 (1.19 to 2.82) \\
\textbf{EW(H$\delta$) [Å]} & \phantom{$-$}5.47 (3.85 to 7.44) & \phantom{$-$}4.40 (3.35 to 6.39) & 3.96 (3.28 to 5.69) \\
\textbf{EW({[}OII{]}$\lambda$3727) [Å]} & $-1.74$ ($-3.19$ to $-0.75$) & $-0.65$ ($-1.59$ to $-0.23$) & 1.94 ($-0.25$ to 5.96) \\
\textbf{EW({[}OII{]}$\lambda$3729) [Å]} & $-2.02$ ($-3.57$ to $-0.90$) & $-0.77$ ($-1.72$ to $-0.25$) & 1.01 ($-1.04$ to 5.84) \\ \bottomrule
\end{tabular}
\tablefoot{16\textsuperscript{th} and 84\textsuperscript{th} percentiles are shown in parentheses. }
\label{tab:group_properties_meusinger}
\end{table*}

\subsection{Morphologies and merger histories}
The merger histories of PSB galaxies give insight into the mechanisms that produce them, including whether the PSB phase is externally or internally triggered. Although the process of merging disturbs the morphology of the involved galaxies, it is challenging to constrain the merger history of a galaxy. This is commonly done either by visual inspection or by determining quantitative morphological parameters that are suggestive of past or ongoing mergers \citep{Pawlik_2015,Lotz_2008}. 

\begin{figure}
    \centering
    \includegraphics[width=\linewidth]{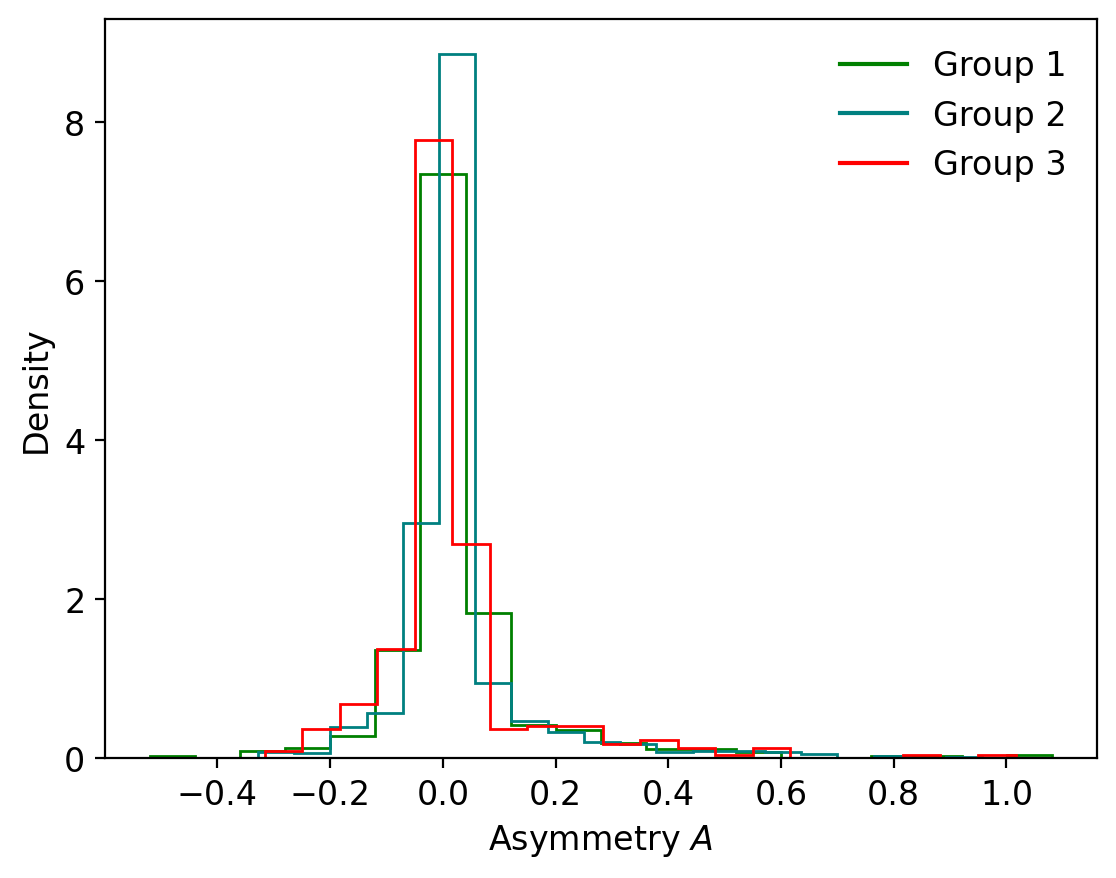}
    \caption{Asymmetry distributions for the three PSB groups in the \citet{Meusinger_2017} catalog. The asymmetry parameter was measured using statmorph, following the method of \citet{Lotz_2004}. }
    \label{fig:asymmetry}
\end{figure}

To examine whether the grouping is correlated to the merger histories of the galaxies, several quantitative morphological parameters were compared between the groups. These include CAS-statistics, Gini/M20, $\chi^2$ of Sersic fits \citep[e.g.,][]{Lotz_2004,Pawlik_2015}, and merger probabilities calculated from visual classifications by nonexperts in Galaxy Zoo \citep{Willett_2013}. The properties are not clearly divided between the groups, but Fig. \ref{fig:asymmetry} shows a variation in the asymmetry parameter, which was measured from galaxy images using statmorph \citep{Rodriguez_Gomez_2018} and quantifies the rotational symmetry of the light from the galaxy \citep{Lotz_2004}. On average, Group 3 is less asymmetric than Groups 1 and 2, and KS tests showed that Group 3 is inconsistent with being identical to Groups 1 and 2 ($p<0.01$). Thus, Group 3 is probably either associated with a minor merger or with no merger at all. Groups 1 and 2 are weakly inconsistent with being identical ($p\sim 0.017$). These differences could be blurred by poor preprocessing of the images.

\citet{Li_2023} presented a different catalog of PSB galaxies that were visually classified as either post-merger or noninteracting (and thereby, nonpost-merger). This was done following the method by \citet{Nair_2010}, where post-mergers are found among galaxies with unusual forms in their g-band images. The \citet{Li_2023} galaxies were selected to have statistically similar stellar masses and redshifts (in the range $0.02\leq z\leq 0.06$). Requiring EW(H$\delta$) $>3$ Å, EW(H$\alpha$) $>-10$ Å, and log($-$EW([OII])) $<0.23\times$ EW(H$\delta$) $-0.46$, the catalog includes 264 resolved PSB galaxies, 96 of which are identified as post-merger. 

Although the \citet{Li_2023} PSB sample is too small to reliably identify any groups in a UMAP embedding, some structures can still be studied. The UMAP embeddings of the \citet{Li_2023} galaxies suggest that the spectral properties of a PSB galaxy are correlated to its merger history: If there were no correlation between the merger history of a galaxy and its spectral properties, the post-mergers should be randomly distributed across the embedding because UMAP is given no information about any morphological features of the galaxies. Thus, the fraction of post-mergers in any sufficiently large section should be similar to that of the entire catalog. However, in a selected region that includes 25\% of the galaxies in the catalog and 50\% of the post-mergers, 71\% are post-merger, as opposed to the expected 36\% if the post-mergers were randomly distributed. This excess factor of two indicates that post-merger PSB galaxies share some spectral properties that separate them from nonpost-merger PSB galaxies. This is consistent with there being multiple types of PSB galaxies and multiple mechanisms for quenching star formation. Because merger histories are difficult to constrain, the post-merger fraction in the selected region might be even larger. Similarly, \citet{Ventou_2019} suggested a merger fraction of $\sim20\%$ for galaxies at redshift $z<1.5$, which is comparable to the merger fraction of 25\% among the remaining galaxies. 

\subsection{Derived properties}
\label{subsec:other_properties}
Additional derived galaxy properties can be used to evaluate potential explanations for the origins of these three groups. In principle, these derived properties could have been used to produce the embedding described in Sect. \ref{sec:identifying_psb_groups}.  They were measured via photometric template fitting, however, and therefore relied on several strong assumptions (e.g., of the initial mass function or the shape of the star formation history). Consequently, only properties in existing catalogs that were derived using established techniques were compared. Further, individual galaxy properties are often only very weakly constrained, so that the most useful results instead come from considering distributions of properties across large galaxy samples \citep{Ilbert_2013,Mitchell2013,Speagle_2014,Davidzon_2017,Weaver_2023}.  Thus, it is meaningful to test whether the groups have similar or dissimilar distributions of the derived properties, even though the individual values may not be robust. 

Perhaps the most informative are the stellar mass distributions. It has been known for several decades that at fixed redshift, the most massive galaxies are predominantly quiescent while the least massive are still forming stars. This result is one of several that are collectively termed downsizing (see \citet{Cowie_1996}). Galaxies with stellar masses similar to the most massive star-forming galaxies will be quiescent at lower redshifts, implying that these galaxies must turn off in the near future.   

\begin{figure}
    \centering
    \includegraphics[width=\linewidth]{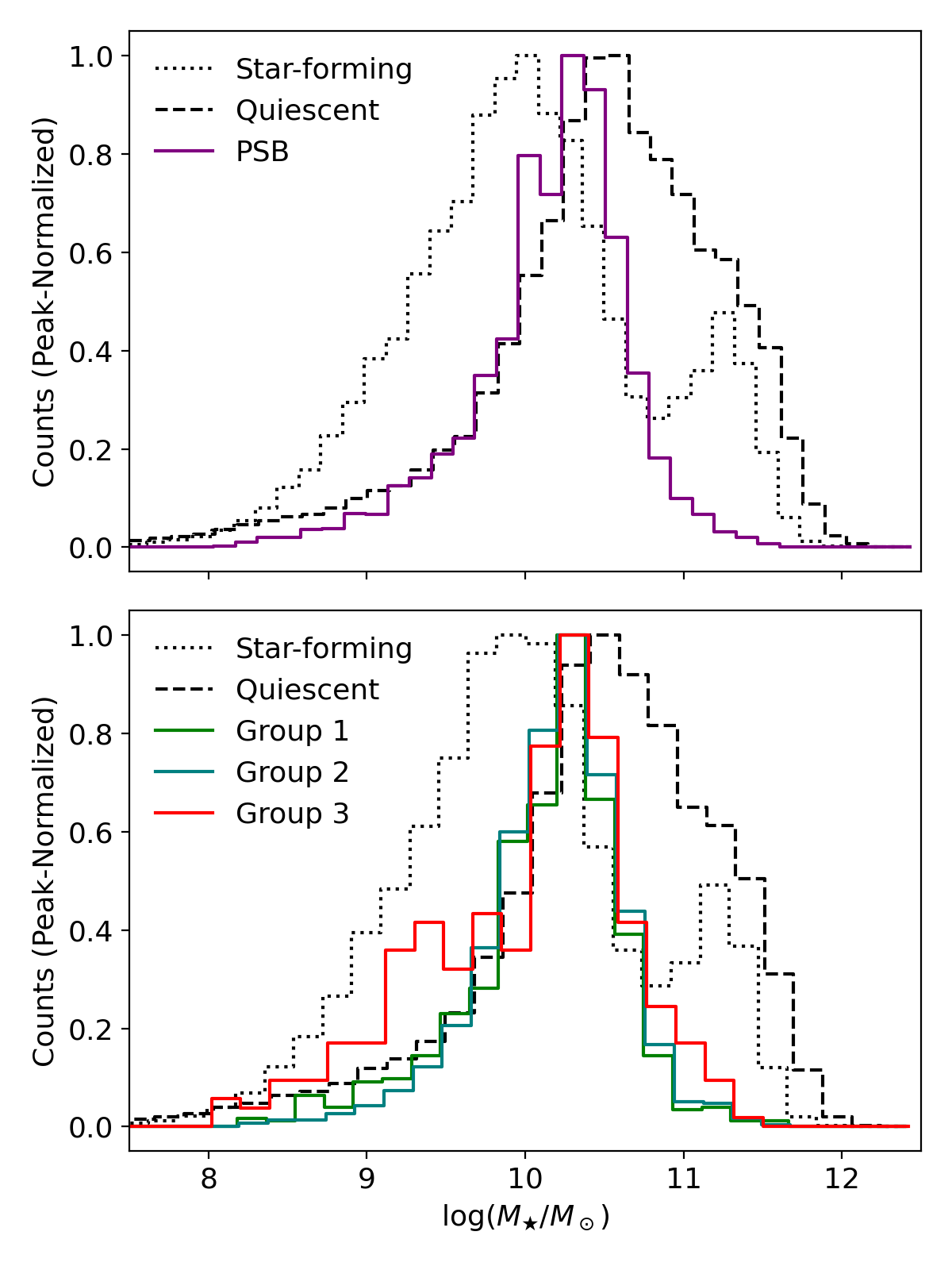}
    \caption{Stellar mass distribution of PSB galaxies in the \citet{Meusinger_2017} catalog compared to star-forming and quiescent galaxies. Stellar masses were taken from the Portsmouth catalog, and the quiescent sample was selected selected to have SFR $=0$. }
    \label{fig:mass_distribution}
\end{figure}

\begin{figure}
    \centering
    \includegraphics[width=.95\linewidth]{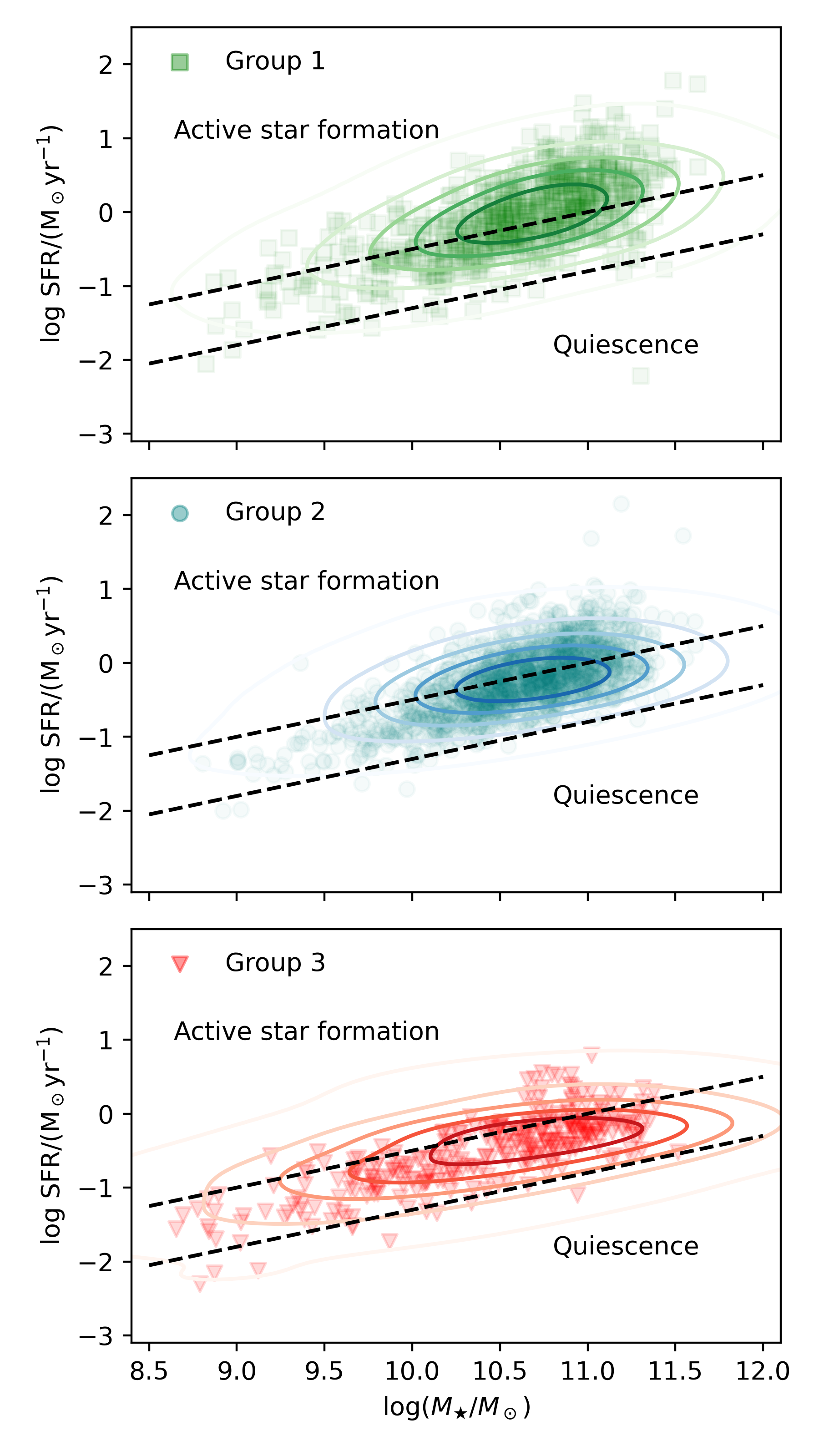}
    \caption{Star formation rate vs. stellar mass parameter space for each of the PSB groups in the \citet{Meusinger_2017} catalog. Groups 1, 2, and 3 are shown from top to bottom. The galaxy properties were taken from the MPA-JHU catalog, and the transitional region is outlined by the dashed lines following \citet{Li_2023}.}
    \label{fig:m_vs_sfr}
\end{figure}

Fig. \ref{fig:mass_distribution} shows that all three groups of PSB galaxies have masses similar to both the most massive star-forming galaxies (SFGs) and the least massive quiescent galaxies (QGs), corresponding to galaxies that should be at or near turnoff. Thus, it would appear that all three groups of PSB galaxies might be associated not with rejuvenation of typical quiescent galaxies, in which case their stellar mass distribution should be more similar to the quiescent stellar mass distribution, but rather with turnoff from the star-forming main sequence. Further, Fig. \ref{fig:m_vs_sfr} shows that galaxies within the three PSB groups migrate from the star-forming region (Group 1) to the transitional region (Group 3) on a SFR vs. stellar mass diagram, suggesting the possibility that the three might comprise an age sequence within a single scenario for main-sequence turnoff. Based on ages from the star-forming model in the Portsmouth catalog, however, the galaxies in all three groups are consistent with having identical age distributions ($p>0.1$).

Fig. \ref{fig:mass_distribution} further shows that Groups 1 and 2 have statistically similar mass distributions ($p>0.1$), whereas Group 3 is differs, predominantly on the low-mass end ($p\ll0.01$). This is true for stellar masses measured by both the MPA-JHU catalog and the Portsmouth catalog. As a result, this deviation is likely genuine, although there is no obvious mechanism that would produce it. One possibility is that Group 3 should actually comprise two groups, and that the limited information provided to UMAP is insufficient to separate them. 

The Portsmouth catalog fits stellar evolution models to SDSS photometry with a small number of discrete choices of metallicity ($Z=0.004,\;0.01,\;0.02$, and $0.04$). All three galaxy groups have broadly similar metallicity distributions that peak between the peaks of the star-forming and quiescent distributions in the $Z=0.02$ bin. A $\chi^2$ test indicates that the small differences in the distributions of the three groups are statistically significant ($p<0.01$), but a spectroscopic study allowing a continuous range of metallicities to be fit is required to properly determine the origin and potential astrophysical significance.

\section{Potential origins of PSB galaxy types}
\label{sec:potential_origins}

In Sect. \ref{sec:identifying_psb_groups} and \ref{sec:psb_properties}, we argued that there are three distinct types of PSB galaxies rather than a single type, as previously proposed. We considered several possibilities to explain the origins of these distinct types:
\begin{enumerate}
    \item The grouping might be an evolutionary sequence in which Group 1 is the youngest and Group 3 is the oldest.
    \item The groups might all be merger-driven, but produced by different types of mergers.
    \item Some of the groups might instead evolve secularly through the PSB phase, such as in the red star-forming galaxy scenario \citep{Steinhardt_2025}. 
\end{enumerate}
A summary of the results is given in Table \ref{tab:overview_model_results}.

\begin{table*}
    \centering
    \caption{Overview of predictions for a selection of proposed models and the observations we used to test them.} 
    \newcolumntype{L}{!{\color{white}\ }l}
    \begin{tabular}{LLLLL}
    \toprule
         \textbf{Measurement} &
         \textbf{Age sequence} &
         \textbf{Merger-driven} &
         \textbf{RSFG prediction} &
         \textbf{Observation} \\
         & 
         \textbf{prediction} &
         \textbf{mechanisms} &
         & \\
         & & \textbf{prediction} & & \\
    \midrule
         Emission line &
         \cellcolor[rgb]{0.851,0.949,0.816}Decrease towards &
         \cellcolor[rgb]{0.851,0.949,0.816}Decrease towards &
         \cellcolor[rgb]{0.851,0.949,0.816}Weak in Group 3 &
         Decrease towards \\
         strengths &
         \cellcolor[rgb]{0.851,0.949,0.816}Group 3 &
         \cellcolor[rgb]{0.851,0.949,0.816}Group 3 &
         \cellcolor[rgb]{0.851,0.949,0.816} & 
         Group 3 \\
         \addlinespace[.25em]
         
         Mass distribution &
         \cellcolor[rgb]{0.961,0.957,0.808}Similar &
         \cellcolor[rgb]{0.961,0.957,0.808}Group 1-2 like SFGs, &
         \cellcolor[rgb]{0.961,0.957,0.808}Could be dissimilar &
         Groups 1-2 similar, \\ 
         &
         \cellcolor[rgb]{0.961,0.957,0.808} &
         \cellcolor[rgb]{0.961,0.957,0.808}Group 3 like QGs &
         \cellcolor[rgb]{0.961,0.957,0.808} &
         Group 3 different \\
         \addlinespace[.25em]
         
         Age distribution &
         \cellcolor[rgb]{0.984,0.89,0.839}Increase towards &
         \cellcolor[rgb]{0.961,0.957,0.808}Higher in Group 3 &
         \cellcolor[rgb]{0.851,0.949,0.816}Could be similar &
         Similar \\
         &
         \cellcolor[rgb]{0.984,0.89,0.839}Group 3 &
         \cellcolor[rgb]{0.961,0.957,0.808} & 
         \cellcolor[rgb]{0.851,0.949,0.816} & \\
         \addlinespace[.25em]

         Extinction &
         \cellcolor[rgb]{0.961,0.957,0.808}Decrease towards &
         &
         &
         Similar \\
         &
         \cellcolor[rgb]{0.961,0.957,0.808}Group 3 &
         & 
         & \\
         \addlinespace[.25em]

         SFR &
         \cellcolor[rgb]{0.851,0.949,0.816}Decrease towards &
         &
         \cellcolor[rgb]{0.851,0.949,0.816}Low in Group 3 &
         Decrease towards \\
         &
         \cellcolor[rgb]{0.851,0.949,0.816}Group 3 &
         &
         \cellcolor[rgb]{0.851,0.949,0.816} &
         Group 3 \\
         \addlinespace[.25em]

         Asymmetry &
         \cellcolor[rgb]{0.851,0.949,0.816}Decrease towards &
         \cellcolor[rgb]{0.851,0.949,0.816}Lower in Group 3 &
         \cellcolor[rgb]{0.851,0.949,0.816}Low in Group 3 &
         Lower in Group 3 \\
         & \cellcolor[rgb]{0.851,0.949,0.816}Group 3 &
         \cellcolor[rgb]{0.851,0.949,0.816} & 
         \cellcolor[rgb]{0.851,0.949,0.816} & \\         
    \bottomrule
    \end{tabular}
    \tablefoot{The green, yellow, and red cells indicates that the model is either consistent, weakly (in)consistent, or inconsistent with the observations, respectively.}
\label{tab:overview_model_results}
\end{table*}

\subsection{An evolutionary sequence}
\label{subsec:evolutionary_sequence}
One interpretation is that the grouping is an evolutionary sequence of PSB galaxies with a similar origin. We assumed that Group 1 is the youngest (and Group 3 the oldest) because H$\alpha$ and [OII] emission and SFRs decrease continuously toward Group 3. This is consistent with the canonical scenario according to which PSB galaxies are in the midst of rapid quenching. Further, Group 1 has greater continuum emission toward UV wavelengths and emits some [OIII], indicating that some of its O and B stars have not yet died. 

If the PSB phase has been initiated by a merger (as is commonly thought), the evidence of the merger will fade over time. If the grouping is an evolutionary sequence, the evidence of the merger should therefore be less pronounced in Group 3. This is consistent with observations, as Group 3 is less asymmetric than Group 1. The partial separation of post-mergers and nonpost-mergers in the \citet{Li_2023} catalog is also consistent with the grouping being an evolutionary sequence of merger-driven PSB galaxies: Obvious post-mergers and nonpost-mergers are easily categorized, whereas recovered post-mergers may be incorrectly categorized as nonpost-mergers. Likewise, the size of the merger may affect the extent to which the morphology of the galaxy is disturbed, causing variations in the asymmetry parameter \citep{Lotz_2008_morph}. 

Observations indicate, however, that the galaxies are equally old, suggesting that the grouping is not an evolutionary sequence. On the other hand, it is difficult to determine the age of a galaxy using photometric template fitting (as in this case; see \citealt{Pforr2012}), and a high resolution is needed to differentiate between the ages of galaxies in a short-lived phase like this. Further, this method is dominated by the continuum, whereas the SEDs primarily differ in the lines (Fig. \ref{fig:stacked_spectra}). In addition, the galaxy age is only a good proxy of the time elapsed since the initiation of the PSB phase if the galaxies on average were equally old when the PSB phase was initiated. If the groups only differ by the evolutionary stage of the PSB phase in which they are, however, there is no obvious reason why the groups on average should have initiated the PSB phase at different ages. 

If the groups comprise an age sequence, they should have nearly identical stellar mass distributions because they have similar origins and none of them form stars actively. The mass distribution of Group 3 is statistically different from those of Groups 1 and 2 (Fig. \ref{fig:mass_distribution}), which would be inconsistent with the age-sequence hypothesis. The discrepancy is limited to a small fraction of Group 3 on the low-mass end, however.  Thus, another possibility is that Group 3 comprises two groups, one at the low-mass end and another with a mass distribution similar to Groups 1 and 2.

\subsection{Distinct merger-driven origins}
\label{subsec:multiple_merger_origins}
As shown in Sect. \ref{sec:psb_properties}, the properties of PSB galaxies are multimodal.  Thus, the three groups likely do not constitute a single evolutionary sequence, but instead have distinct astrophysical origins. The PSB phase might still be entirely merger-driven, however. 

As described in Sect. \ref{sec:intro}, \citet{Pawlik_2019} presented three distinct merger-driven mechanisms for producing PSB galaxies, differentiated by the evolutionary stage of a galaxy when the merger occurs. The three groups might then be caused by mergers creating a starburst and subsequent post-starburst phase in star-forming galaxies, turnoff galaxies, and quiescent galaxies.  

Since Group 1 has the highest SFR and the strongest emission lines, this group may represent the \citet{Pawlik_2019} blue-to-blue cycle in which the galaxies resume their star formation after the PSB phase. Group 2 may include galaxies that transition from star-forming to quiescent, and the low asymmetry of Group 3 suggests that its (formerly quiescent) galaxies may have been rejuvenated by minor mergers. Moreover, Group 3 should have weaker emission lines because its star formation has already quenched, and the minor merger has only caused a weak starburst. This is consistent with the general trend observed in Fig. \ref{fig:m_vs_sfr}. 

If the groups are produced by mergers affecting different types of precursor galaxies, the mass distribution of Groups 1 and 2 (originating from star-forming galaxies) should be similar to star-forming galaxies, whereas the mass distribution of Group 3 (originating from quiescent galaxies) should be similar to that of quiescent galaxies. This prediction is consistent with observations because the masses of all three PSB groups are similar to the most massive star-forming galaxies and to the least massive quiescent galaxies (see Fig. \ref{fig:mass_distribution}). 

\citet{Pawlik_2019} further suggested that the blue-to-blue cycle occurs for $\sim30\%$ of PSB galaxies with $9.5<\log(M_*/M_\odot)<10.5$, which is consistent with Group 1 containing $\sim30\%$ of the galaxies within this mass range. \citet{Pawlik_2019} also suggested that $\sim70\%$ of PSB galaxies with $9.5<\log(M_*/M_\odot)<10.5$ and $\sim60\%$ of PSB galaxies with $\log(M_*/M_\odot)>10.5$ transition from star-forming to quiescent as a result of a gas-rich major merger, which is consistent with Group 2 containing $\sim60\%$ of the galaxies within both of these mass ranges. \citet{Pawlik_2019} suggested, however, that $\sim40\%$ of PSB galaxies with $\log(M_*/M_\odot)>10.5$ are rejuvenated galaxies, whereas Group 3 only contains $\sim10\%$ of the \citet{Meusinger_2017} galaxies within this mass range. \citet{Pawlik_2019} also suggested that these galaxies gradually move toward the high-mass end of the red sequence, but Group 3 is found at the low-mass end. In contrast to observations, in this scenario, Group 3 is also expected to have an older stellar population because it was previously quiescent. Instead, the stellar populations in all three groups have similar age distributions. 

\subsection{RSFGs and secular evolution}
\label{subsec:rsfg}
The \citet{Pawlik_2019} interpretation might be consistent with observations of Groups 1 and 2, but is apparently inconsistent with Group 3. A plausible interpretation must be able to explain why the galaxies in Group 3 are the reddest, but are younger than quiescent galaxies and lie at the low-mass end of the red sequence. 

One possibility is that Group 3 contains the nonmerger-driven but secularly evolving red star-forming galaxies proposed by \citet{Steinhardt_2025}. This interpretation is consistent with Group 3 having weaker emission lines because RSFGs have stopped to actively form O and B stars. The model also explains why Group 3 has a lower SFR because SFRs are traced by short-lived O and B stars, not by the low-mass stars that RSFGs are still actively forming. If Group 3 is produced by a different mechanism, this might even explain why Group 3 is most clearly separated in the UMAP embedding. Moreover, the enhanced low-wavelength continuum emission of Group 3 compared to Group 2 is consistent with Group 3 being produced by a different mechanism, rather than being the end stage of an evolutionary sequence.

In addition, all three galaxy types could be equally old (as observations suggest), with some evolving secularly and some being forced into the PSB-phase by mergers. The nonmerger-driven Group 3 will have a lower asymmetry, as observed. 

Finally, the mass distribution of Group 3 (ranging from the most massive star-forming galaxies to the least massive quiescent ones) is consistent with RSFGs being at turnoff. Further, Group 3 could (but is not required to) have a different mass distribution if it is produced by a different mechanism, which is consistent with its deviation from Groups 1 and 2 at the low-mass end. \citet{Steinhardt_2025} made no additional predictions for the mass distribution of RSFGs compared to merger-driven PSB galaxies.

\section{Discussion}
\label{sec:discussion}
We used UMAP to cluster PSB galaxies based on the strengths of selected spectral lines and found strong evidence that there are three distinct types of post-starburst galaxies with dissimilar properties. It appears that the grouping is not simply an age sequence, but is instead correlated with additional observed galaxy properties and likely also with the merger histories of these galaxies. 

The most likely scenario is that the three types of PSB galaxies have distinct astrophysical origins, some of which may not be associated with a merger or a starburst. Groups 1 and 2 have several properties in common with merger-driven scenarios, but Group 3 appears to be inconsistent.  The hypothesis that galaxies might not quench when they leave the main sequence, but rather continue as red star-forming galaxies (RSFG; \citealt{Steinhardt_2025}), is consistent with observations of Group 3.

Additional falsifiable tests are needed to determine the exact origins of the three PSB galaxy types, however. Moreover (for the properties compared in this work), the RSFG hypothesis makes less specific predictions that are harder to reject. Thus, future research could compare the quenching directions of the PSB galaxy types to decide whether the PSB phase(s) is (are) internally or externally triggered: \citet{Li_2023} reported that post-mergers predominantly quench outside-in, whereas RSFGs are expected to quench inside-out \citep{Steinhardt_2025}. If there is no correlation between the PSB groups and their quenching directions, the RSFG interpretation must therefore be rejected. 

Interestingly, the mass distributions of all three PSB groups range from the most massive star-forming galaxies to the least massive quenched galaxies. This is expected for Group 3 if it contains RSFGs that are at turnoff; that all three PSB galaxy types have masses in this range indicates that they are all near turnoff. This observation is consistent with downsizing, the idea that the most massive galaxies formed the bulk of their stellar mass on a shorter timescale. Similarly, \citet{Wong_2012} suggested that PSB galaxies contribute to building up the low-mass end of the red sequence. This is interesting because PSB galaxies are thought to be created by some rare event, but if they are all near turnoff, maybe all typical galaxies go through the post-starburst phase. 

Generally, downsizing is inconsistent with any interpretation in which the galaxies do not transition from star-forming to quiescent, for example, the rejuvenation of galaxies or the blue-to-blue cycle interpretation of Group 1. The hypothesis does not constrain the quenching mechanism(s) further. Thus, other interpretations may be consistent with downsizing as well. There would be no reason for post-mergers to be over-represented among PSB galaxies if the majority of galaxies evolved secularly through the PSB phase, however. Therefore, at least one of the PSB groups is likely to be associated with merging. 

\begin{acknowledgements}
    The authors would like to thank Eva Gregersen, Preethi Nair, Jens Nielsen, Tom Reynolds and Darach Watson for helpful comments. EN, CS, MH, and AS were supported by research grants (VIL16599, VIL54489) from VILLUM FONDEN.
\end{acknowledgements}

\bibliographystyle{aa}
\bibliography{refs.bib} 

\begin{thebibliography}{35}
\expandafter\ifx\csname natexlab\endcsname\relax\def\natexlab#1{#1}\fi

\bibitem[{Abazajian {et~al.}(2009)Abazajian, Adelman-McCarthy, Agüeros, Allam, Prieto, An, Anderson, Anderson, Annis, Bahcall, Bailer-Jones, Barentine, Bassett, Becker, Beers, Bell, Belokurov, Berlind, Berman, Bernardi, Bickerton, Bizyaev, Blakeslee, Blanton, Bochanski, Boroski, Brewington, Brinchmann, Brinkmann, Brunner, Budavári, Carey, Carliles, Carr, Castander, Cinabro, Connolly, Csabai, Cunha, Czarapata, Davenport, de~Haas, Dilday, Doi, Eisenstein, Evans, Evans, Fan, Friedman, Frieman, Fukugita, Gänsicke, Gates, Gillespie, Gilmore, Gonzalez, Gonzalez, Grebel, Gunn, Györy, Hall, Harding, Harris, Harvanek, Hawley, Hayes, Heckman, Hendry, Hennessy, Hindsley, Hoblitt, Hogan, Hogg, Holtzman, Hyde, Ichikawa, Ichikawa, Im, Ivezić, Jester, Jiang, Johnson, Jorgensen, Jurić, Kent, Kessler, Kleinman, Knapp, Konishi, Kron, Krzesinski, Kuropatkin, Lampeitl, Lebedeva, Lee, Lee, Leger, Lépine, Li, Lima, Lin, Long, Loomis, Loveday, Lupton, Magnier, Malanushenko, Malanushenko, Mandelbaum, Margon, Marriner,
  Martínez-Delgado, Matsubara, McGehee, McKay, Meiksin, Morrison, Mullally, Munn, Murphy, Nash, Nebot, Neilsen, Newberg, Newman, Nichol, Nicinski, Nieto-Santisteban, Nitta, Okamura, Oravetz, Ostriker, Owen, Padmanabhan, Pan, Park, Pauls, Peoples, Percival, Pier, Pope, Pourbaix, Price, Purger, Quinn, Raddick, Fiorentin, Richards, Richmond, Riess, Rix, Rockosi, Sako, Schlegel, Schneider, Scholz, Schreiber, Schwope, Seljak, Sesar, Sheldon, Shimasaku, Sibley, Simmons, Sivarani, Smith, Smith, Smolčić, Snedden, Stebbins, Steinmetz, Stoughton, Strauss, SubbaRao, Suto, Szalay, Szapudi, Szkody, Tanaka, Tegmark, Teodoro, Thakar, Tremonti, Tucker, Uomoto, Vanden~Berk, Vandenberg, Vidrih, Vogeley, Voges, Vogt, Wadadekar, Watters, Weinberg, West, White, Wilhite, Wonders, Yanny, Yocum, York, Zehavi, Zibetti, \& Zucker}]{Abazajian_2009}
Abazajian, K.~N., Adelman-McCarthy, J.~K., Agüeros, M.~A., {et~al.} 2009, \apjs, 182, 543–558

\bibitem[{Baldwin {et~al.}(1981)Baldwin, Phillips, \& Terlevich}]{Baldwin_1981}
Baldwin, J.~A., Phillips, M.~M., \& Terlevich, R. 1981, \pasp, 93, 5

\bibitem[{Brinchmann {et~al.}(2004)Brinchmann, Charlot, White, Tremonti, Kauffmann, Heckman, \& Brinkmann}]{Brinchmann_2004}
Brinchmann, J., Charlot, S., White, S. D.~M., {et~al.} 2004, \mnras, 351, 1151–1179

\bibitem[{Chambers {et~al.}(2019)Chambers, Magnier, Metcalfe, Flewelling, Huber, Waters, Denneau, Draper, Farrow, Finkbeiner, Holmberg, Koppenhoefer, Price, Rest, Saglia, Schlafly, Smartt, Sweeney, Wainscoat, Burgett, Chastel, Grav, Heasley, Hodapp, Jedicke, Kaiser, Kudritzki, Luppino, Lupton, Monet, Morgan, Onaka, Shiao, Stubbs, Tonry, White, Bañados, Bell, Bender, Bernard, Boegner, Boffi, Botticella, Calamida, Casertano, Chen, Chen, Cole, Deacon, Frenk, Fitzsimmons, Gezari, Gibbs, Goessl, Goggia, Gourgue, Goldman, Grant, Grebel, Hambly, Hasinger, Heavens, Heckman, Henderson, Henning, Holman, Hopp, Ip, Isani, Jackson, Keyes, Koekemoer, Kotak, Le, Liska, Long, Lucey, Liu, Martin, Masci, McLean, Mindel, Misra, Morganson, Murphy, Obaika, Narayan, Nieto-Santisteban, Norberg, Peacock, Pier, Postman, Primak, Rae, Rai, Riess, Riffeser, Rix, Röser, Russel, Rutz, Schilbach, Schultz, Scolnic, Strolger, Szalay, Seitz, Small, Smith, Soderblom, Taylor, Thomson, Taylor, Thakar, Thiel, Thilker, Unger, Urata, Valenti,
  Wagner, Walder, Walter, Watters, Werner, Wood-Vasey, \& Wyse}]{chambers2019panstarrs1surveys}
Chambers, K.~C., Magnier, E.~A., Metcalfe, N., {et~al.} 2019, The Pan-STARRS1 Surveys

\bibitem[{Cheng {et~al.}(2024)Cheng, Li, Li, Yan, \& Mo}]{Cheng_2024}
Cheng, Z., Li, C., Li, N., Yan, R., \& Mo, H. 2024, \apj, 961, 216

\bibitem[{Cowie {et~al.}(1996)Cowie, Songaila, Hu, \& Cohen}]{Cowie_1996}
Cowie, L.~L., Songaila, A., Hu, E.~M., \& Cohen, J.~G. 1996, \apj, 112, 839

\bibitem[{Davidzon {et~al.}(2017)Davidzon, Ilbert, Laigle, Coupon, McCracken, Delvecchio, Masters, Capak, Hsieh, Le~Fèvre, Tresse, Bethermin, Chang, Faisst, Le~Floc’h, Steinhardt, Toft, Aussel, Dubois, Hasinger, Salvato, Sanders, Scoville, \& Silverman}]{Davidzon_2017}
Davidzon, I., Ilbert, O., Laigle, C., {et~al.} 2017, \aap, 605, A70

\bibitem[{French(2021)}]{French_2021}
French, K.~D. 2021, PASP, 133, 072001

\bibitem[{Ilbert {et~al.}(2013)Ilbert, McCracken, Le~Fèvre, Capak, Dunlop, Karim, Renzini, Caputi, Boissier, Arnouts, Aussel, Comparat, Guo, Hudelot, Kartaltepe, Kneib, Krogager, Le~Floc’h, Lilly, Mellier, Milvang-Jensen, Moutard, Onodera, Richard, Salvato, Sanders, Scoville, Silverman, Taniguchi, Tasca, Thomas, Toft, Tresse, Vergani, Wolk, \& Zirm}]{Ilbert_2013}
Ilbert, O., McCracken, H.~J., Le~Fèvre, O., {et~al.} 2013, \aap, 556, A55

\bibitem[{Kauffmann {et~al.}(2003{\natexlab{a}})Kauffmann, Heckman, Simon~White, Charlot, Tremonti, Brinchmann, Bruzual, Peng, Seibert, Bernardi, Blanton, Brinkmann, Castander, Csábai, Fukugita, Ivezic, Munn, Nichol, Padmanabhan, Thakar, Weinberg, \& York}]{Kauffmann_2003}
Kauffmann, G., Heckman, T.~M., Simon~White, D.~M., {et~al.} 2003{\natexlab{a}}, \mnras, 341, 33–53

\bibitem[{Kauffmann {et~al.}(2003{\natexlab{b}})Kauffmann, Heckman, Tremonti, Brinchmann, Charlot, White, Ridgway, Brinkmann, Fukugita, Hall, Ivezić, Richards, \& Schneider}]{Kauffmann_2003_agn}
Kauffmann, G., Heckman, T.~M., Tremonti, C., {et~al.} 2003{\natexlab{b}}, \mnras, 346, 1055–1077

\bibitem[{Kewley {et~al.}(2013)Kewley, Maier, Yabe, Ohta, Akiyama, Dopita, \& Yuan}]{Kewley_2013}
Kewley, L.~J., Maier, C., Yabe, K., {et~al.} 2013, \apj, 774, L10

\bibitem[{Li {et~al.}(2023)Li, Nair, Rowlands, Masters, Stark, Drory, Ellison, Irwin, Satyapal, Jones, Keel, Mukundan, \& Tu}]{Li_2023}
Li, W., Nair, P., Rowlands, K., {et~al.} 2023, \mnras, 523, 720–738

\bibitem[{Lotz {et~al.}(2008{\natexlab{a}})Lotz, Davis, Faber, Guhathakurta, Gwyn, Huang, Koo, Le~Floc’h, Lin, Newman, Noeske, Papovich, Willmer, Coil, Conselice, Cooper, Hopkins, Metevier, Primack, Rieke, \& Weiner}]{Lotz_2008}
Lotz, J.~M., Davis, M., Faber, S.~M., {et~al.} 2008{\natexlab{a}}, \apj, 672, 177–197

\bibitem[{Lotz {et~al.}(2008{\natexlab{b}})Lotz, Jonsson, Cox, \& Primack}]{Lotz_2008_morph}
Lotz, J.~M., Jonsson, P., Cox, T.~J., \& Primack, J.~R. 2008{\natexlab{b}}, \mnras, 391, 1137–1162

\bibitem[{Lotz {et~al.}(2004)Lotz, Primack, \& Madau}]{Lotz_2004}
Lotz, J.~M., Primack, J., \& Madau, P. 2004, \apj, 128, 163–182

\bibitem[{Maraston {et~al.}(2013)Maraston, Pforr, Henriques, Thomas, Wake, Brownstein, Capozzi, Tinker, Bundy, Skibba, Beifiori, Nichol, Edmondson, Schneider, Chen, Masters, Steele, Bolton, York, Weaver, Higgs, Bizyaev, Brewington, Malanushenko, Malanushenko, Snedden, Oravetz, Pan, Shelden, \& Simmons}]{Maraston_2013}
Maraston, C., Pforr, J., Henriques, B.~M., {et~al.} 2013, \mnras, 435, 2764–2792

\bibitem[{McInnes {et~al.}(2020)McInnes, Healy, \& Melville}]{mcinnes2020umap}
McInnes, L., Healy, J., \& Melville, J. 2020, UMAP: Uniform Manifold Approximation and Projection for Dimension Reduction

\bibitem[{Meusinger {et~al.}(2017)Meusinger, Brünecke, Schalldach, \& in~der Au}]{Meusinger_2017}
Meusinger, H., Brünecke, J., Schalldach, P., \& in~der Au, A. 2017, \aap, 597, A134

\bibitem[{{Mitchell} {et~al.}(2013){Mitchell}, {Lacey}, {Baugh}, \& {Cole}}]{Mitchell2013}
{Mitchell}, P.~D., {Lacey}, C.~G., {Baugh}, C.~M., \& {Cole}, S. 2013, \mnras, 435, 87

\bibitem[{Nair \& Abraham(2010)}]{Nair_2010}
Nair, P.~B. \& Abraham, R.~G. 2010, \apjs, 186, 427–456

\bibitem[{Opitz(2024)}]{Opitz_2024}
Opitz, J. 2024, TACL, 12, 820–836

\bibitem[{Pawlik {et~al.}(2019)Pawlik, McAlpine, Trayford, Wild, Bower, Crain, Schaller, \& Schaye}]{Pawlik_2019}
Pawlik, M.~M., McAlpine, S., Trayford, J.~W., {et~al.} 2019, NA, 3, 440–446

\bibitem[{Pawlik {et~al.}(2015)Pawlik, Wild, Walcher, Johansson, Villforth, Rowlands, Mendez-Abreu, \& Hewlett}]{Pawlik_2015}
Pawlik, M.~M., Wild, V., Walcher, C.~J., {et~al.} 2015, \mnras, 456, 3032–3052

\bibitem[{{Pforr} {et~al.}(2012){Pforr}, {Maraston}, \& {Tonini}}]{Pforr2012}
{Pforr}, J., {Maraston}, C., \& {Tonini}, C. 2012, \mnras, 422, 3285

\bibitem[{Rodriguez-Gomez {et~al.}(2018)Rodriguez-Gomez, Snyder, Lotz, Nelson, Pillepich, Springel, Genel, Weinberger, Tacchella, Pakmor, Torrey, Marinacci, Vogelsberger, Hernquist, \& Thilker}]{Rodriguez_Gomez_2018}
Rodriguez-Gomez, V., Snyder, G.~F., Lotz, J.~M., {et~al.} 2018, \mnras, 483, 4140–4159

\bibitem[{Speagle {et~al.}(2014)Speagle, Steinhardt, Capak, \& Silverman}]{Speagle_2014}
Speagle, J.~S., Steinhardt, C.~L., Capak, P.~L., \& Silverman, J.~D. 2014, \apjs, 214, 15

\bibitem[{Steinhardt(2025)}]{Steinhardt_2025}
Steinhardt, C.~L. 2025, \apj, 982, 189

\bibitem[{{Steinhardt} {et~al.}(2023){Steinhardt}, {Mann}, {Rusakov}, \& {Jespersen}}]{Steinhardt2023}
{Steinhardt}, C.~L., {Mann}, W.~J., {Rusakov}, V., \& {Jespersen}, C.~K. 2023, \apj, 945, 67

\bibitem[{Ventou {et~al.}(2019)Ventou, Contini, Bouché, Epinat, Brinchmann, Inami, Richard, Schroetter, Soucail, Steinmetz, \& Weilbacher}]{Ventou_2019}
Ventou, E., Contini, T., Bouché, N., {et~al.} 2019, \aap, 631, A87

\bibitem[{Vergani {et~al.}(2009)Vergani, Zamorani, Lilly, Lamareille, Halliday, Scodeggio, Vignali, Ciliegi, Bolzonella, Bondi, Kovač, Knobel, Zucca, Caputi, Pozzetti, Bardelli, Mignoli, Iovino, Carollo, Contini, Kneib, Le~Fèvre, Mainieri, Renzini, Bongiorno, Coppa, Cucciati, de~la Torre, de~Ravel, Franzetti, Garilli, Kampczyk, Le~Borgne, Le~Brun, Maier, Pello, Peng, Perez~Montero, Ricciardelli, Silverman, Tanaka, Tasca, Tresse, Abbas, Bottini, Cappi, Cassata, Cimatti, Guzzo, Koekemoer, Leauthaud, Maccagni, Marinoni, McCracken, Memeo, Meneux, Oesch, Porciani, Scaramella, Capak, Sanders, Scoville, \& Taniguchi}]{Vergani_2009}
Vergani, D., Zamorani, G., Lilly, S., {et~al.} 2009, \aap, 509, A42

\bibitem[{Weaver {et~al.}(2023)Weaver, Davidzon, Toft, Ilbert, McCracken, Gould, Jespersen, Steinhardt, Lagos, Capak, Casey, Chartab, Faisst, Hayward, Kartaltepe, Kauffmann, Koekemoer, Kokorev, Laigle, Liu, Long, Magdis, McPartland, Milvang-Jensen, Mobasher, Moneti, Peng, Sanders, Shuntov, Sneppen, Valentino, Zalesky, \& Zamorani}]{Weaver_2023}
Weaver, J.~R., Davidzon, I., Toft, S., {et~al.} 2023, \aap, 677, A184

\bibitem[{Wilkinson {et~al.}(2017)Wilkinson, Pimbblet, \& Stott}]{Wilkinson_2017}
Wilkinson, C.~L., Pimbblet, K.~A., \& Stott, J.~P. 2017, \mnras, 472, 1447–1457

\bibitem[{Willett {et~al.}(2013)Willett, Lintott, Bamford, Masters, Simmons, Casteels, Edmondson, Fortson, Kaviraj, Keel, Melvin, Nichol, Raddick, Schawinski, Simpson, Skibba, Smith, \& Thomas}]{Willett_2013}
Willett, K.~W., Lintott, C.~J., Bamford, S.~P., {et~al.} 2013, \mnras, 435, 2835–2860

\bibitem[{Wong {et~al.}(2012)Wong, Schawinski, Kaviraj, Masters, Nichol, Lintott, Keel, Darg, Bamford, Andreescu, Murray, Raddick, Szalay, Thomas, \& VandenBerg}]{Wong_2012}
Wong, O.~I., Schawinski, K., Kaviraj, S., {et~al.} 2012, \mnras, 420, 1684–1692

\end{thebibliography}

\label{lastpage}
\end{document}